# Reduced-Latency SC Polar Decoder Architectures

Chuan Zhang, Bo Yuan, and Keshab K. Parhi
Department of Electrical and Computer Engineering
University of Minnesota, Twin Cities
Minneapolis, MN 55455, USA
{zhan0884, yuan0103, parhi}@umn.edu

*Abstract*—Polar codes have become one of the most favorable capacity achieving error correction codes (ECC) along with their simple encoding method. However, among the very few prior successive cancellation (SC) polar decoder designs, the required long code length makes the decoding latency high. In this paper, conventional decoding algorithm is transformed with look-ahead techniques. This reduces the decoding latency by 50%. With pipelining and parallel processing schemes, a parallel SC polar decoder is proposed. Sub-structure sharing approach is employed to design the *merged processing element* (PE). Moreover, inspired by the real FFT architecture, this paper presents a novel *input generating circuit* (ICG) block that can generate additional input signals for *merged* PEs on-the-fly. Gate-level analysis has demonstrated that the proposed design shows advantages of 50% decoding latency and twice throughput over the conventional one with similar hardware cost.

## I. INTRODUCTION

Proposed by Arıkan [1], polar codes have been considered as the first "low complexity" scheme which provably achieves the capacity for a fairly wide array of channels. However, most related research is focused on code performance rather than efficient decoder design. Among the few literatures on the latter topic, [1] proposed a straightforward implementation with successive cancellation (SC) algorithm, whose complexity is $\mathcal{O}(N\log_2 N)$. Compared with the belief propagation (BP) algorithm [1]-[2], SC approach is more suitable for hardware design due to its lower complexity. Several further revised SC polar decoders with complexity of $\mathcal{O}(N)$ were presented by [3].

For these conventional polar decoders, decoding a code of length $N$ requires $2(N-1)$ clock cycles. Since modern communication systems require a code length greater than $2^{10}$ is required, the resulting decoding delay is high. Also, restricted by the successive schedule, the highest hardware efficiency of an active stage can only be 50%. In order to achieve faster decoding and higher efficiency, the loop computation is reformulated with look-ahead techniques, which pre-calculate all possible values of the next code bit and then select the correct one with a multiplexer. This paper proposes a nice recursive time chart construction method which succeeds in reducing the decoding latency by 50%. A parallel decoder example is proposed at gate-level with VLSI-DSP techniques. A hardware-efficient *merged processing element* (PE) and the *input generating circuit* (ICG) block, which works best with the whole decoder, are also employed. Comparison results have shown that the proposed design can achieve only half decoding latency and twice higher throughput while maintaining comparable hardware complexity as the conventional one.

The remainder of this paper is organized as follows. A brief review of min-sum SC decoding algorithm is provided in Section II. In Section III, the systematic algorithm to construct the look-ahead scheduling scheme is given in a recursive manner. The parallel polar decoder architectures with gate-level design details are proposed Section IV. The performance estimation and comparison with the state-of-the-art design are presented in Section V. Section VI concludes the paper.

## II. REVIEW OF MIN-SUM SC ALGORITHM

Consider an arbitrary polar code with parameter ($N$, $K$, $\mathcal{A}$, $u_{\mathcal{A}^c}$) [1]. We denote the input vector as $u_1^N$, which consists of a random part $u_{\mathcal{A}}$ and a frozen part $u_{\mathcal{A}^c}$. The corresponding output vector through channel $W_N$ is $y_1^N$ with conditional probability $W_N(y_1^N | u_1^N)$. Define the likelihood ratio (LR) as,

$$L_N^{(i)}(y_1^N, \hat{u}_1^{i-1}) \triangleq \frac{W_N^{(i)}(y_1^N, \hat{u}_1^{i-1} | 0)}{W_N^{(i)}(y_1^N, \hat{u}_1^{i-1} | 1)}. \qquad (1)$$

The a posteriori decision scheme is given as follows. Here $\mathcal{A}^c$ denotes the index set of channels associated with frozen bits. LRs with even and odd indices can be generated by recursively applying Eq. (2) and (3), respectively.

| A Posteriori Decision Scheme with Frozen Bits |
|---|
| 1: **if** $i \in \mathcal{A}^c$ **then** $\hat{u}_i = u_i$; |
| 2: **else** |
| 3:     **if** $L_N^{(i)}(y_1^N, \hat{u}_1^{i-1}) \geq 1$ **then** $\hat{u}_i = 0$; |
| 4:     **else** $\hat{u}_i = 1$; |
| 5:     **endif** |
| 6: **endif** |

$$\begin{aligned}&L_N^{(2i)}(y_1^N, \hat{u}_1^{2i-1}) \\ &= [L_{N/2}^{(i)}(y_1^{N/2}, \hat{u}_{1,o}^{2i-2} \oplus \hat{u}_{1,e}^{2i-2})]^{1-2\hat{u}_{2i-1}} \cdot L_{N/2}^{(i)}(y_{N/2+1}^N, \hat{u}_{1,e}^{2i-2}),\end{aligned} \qquad (2)$$

$$\begin{aligned}&L_N^{(2i-1)}(y_1^N, \hat{u}_1^{2i-2}) \\ &= \frac{L_{N/2}^{(i)}(y_1^{N/2}, \hat{u}_{1,o}^{2i-2} \oplus \hat{u}_{1,e}^{2i-2}) L_{N/2}^{(i)}(y_{N/2+1}^N, \hat{u}_{1,e}^{2i-2}) + 1}{L_{N/2}^{(i)}(y_1^{N/2}, \hat{u}_{1,o}^{2i-2} \oplus \hat{u}_{1,e}^{2i-2}) + L_{N/2}^{(i)}(y_{N/2+1}^N, \hat{u}_{1,e}^{2i-2})}.\end{aligned} \qquad (3)$$

The decoding procedure of a polar code example with $N = 8$ is illustrated in Fig. 1, where Type I and Type II PEs are in charge of Eq. (2) and (3), respectively. The label attached to each PE indicates the clock cycle index when it is activated.

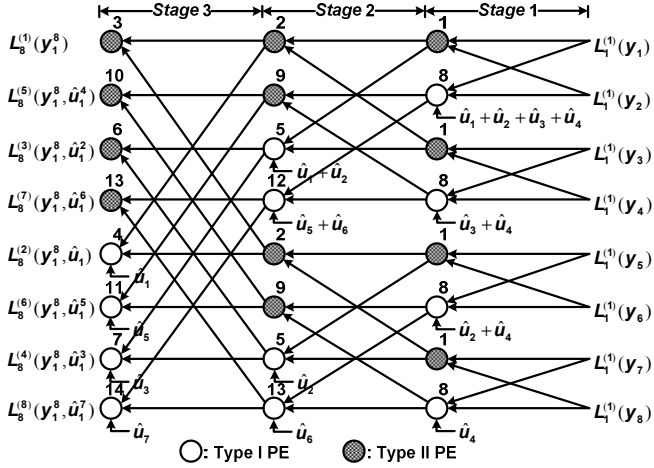

Figure 1: SC decoding process of polar codes with length $N = 8$.

In logarithm domain, Eq. (2) and (3) can be rewritten as:

$$\mathbb{L}_N^{(2i)}(y_1^N, \hat{u}_1^{2i-1}) = (-1)^{\hat{u}_{2i-1}} \mathbb{L}_{N/2}^{(i)}(y_1^{N/2}, \hat{u}_{1,o}^{2i-2} \oplus \hat{u}_{1,e}^{2i-2}) + \mathbb{L}_{N/2}^{(i)}(y_{N/2+1}^N, \hat{u}_{1,e}^{2i-2}), \quad (4)$$

$$\begin{aligned}\mathbb{L}_N^{(2i-1)}(y_1^N, \hat{u}_1^{2i-1}) \\ = 2\operatorname{artanh}\{\tanh[\mathbb{L}_{N/2}^{(i)}(y_1^{N/2}, \hat{u}_{1,o}^{2i-2} \oplus \hat{u}_{1,e}^{2i-2})] \cdot \\ \tanh[\mathbb{L}_{N/2}^{(i)}(y_{N/2+1}^N, \hat{u}_{1,e}^{2i-2})]\}.\end{aligned} \quad (5)$$

$$\mathbb{L}_N^{(i)}(y_1^N, \hat{u}_1^{i-1}) \triangleq \ln L_N^{(i)}(y_1^N, \hat{u}_1^{i-1}). \quad (6)$$

Since large size of look-up table (LUT) is required to implement Eq. (5), it is reduced to the min-sum update rule with sub-optimal approximation:

$$\begin{aligned}\mathbb{L}_N^{(2i-1)}(y_1^N, \hat{u}_1^{2i-2}) \\ \simeq \operatorname{sgn}[\mathbb{L}_{N/2}^{(i)}(y_1^{N/2}, \hat{u}_{1,o}^{2i-2} \oplus \hat{u}_{1,e}^{2i-2})]\operatorname{sgn}[\mathbb{L}_{N/2}^{(i)}(y_{N/2+1}^N, \hat{u}_{1,e}^{2i-2})] \cdot \\ \min[|\mathbb{L}_{N/2}^{(i)}(y_1^{N/2}, \hat{u}_{1,o}^{2i-2} \oplus \hat{u}_{1,e}^{2i-2})|, |\mathbb{L}_{N/2}^{(i)}(y_{N/2+1}^N, \hat{u}_{1,e}^{2i-2})|].\end{aligned} \quad (7)$$

Simulation results have shown that the min-sum SC algorithm, which is LUT free, can keep a balance between decoding performance and hardware efficiency [3], which is very attractive for VLSI designers. Therefore, in the following sections only min-sum SC decoding algorithm is considered.

### III. LATENCY-REDUCED UPDATING SCHEME

However, among all pre-stated algorithms, probabilities are updated according to the same data flow illustrated in Fig. 1, which is straightforward but not efficient. In this section, a high-performance scheme for polar decoder, which only needs half number of clock cycles to obtain the estimated information bits, is developed in a recursive manner. Thorough investigation has revealed that time chart of the straightforward SC decoding process for $N$-bit polar codes can be constructed in recursive way as follows,

**Recursive Construction of Conventional Time Chart**

**1: initializtion** TC= *Null*;

**2: for** $i = \log_2 N, i--,1$ **do**

**3:**   $j = \log_2 N - i + 1$;

**4:**   TC = $\{[j \text{ of Type I}, \text{TC}], i\}$;

**5:**   TC = [TC, TC];

**6:**   change the leftmost $j$ of Type I with $j$ of Type II;

**7: endfor**

**8: output** TC.

Here notation TC = $\{[\mathcal{C}, \text{TC}], s\}$ is used to denote the left insertion of an array $\mathcal{C}$ into the previously arranged time chart TC at *Stage s*. Similarly, TC = [TC, TC] simply means duplicating the previous time chart to obtain the new one. $i$ and $j$ are iterative execution indices. "$j$ of Type I" is the short for "$j$ copy/copies of Type I PE(s) is/are activated in that clock cycle". The corresponding time chart is illustrated in Fig. 2 (a). Since *Stage i* is activated $2^i$ times during the whole decoding process, the total number of clock cycles required is,

$$2\sum_{i=0}^{\log_2 N - 1} 2^i = 2 \cdot \frac{(2^{\log_2 N} - 1)}{2 - 1} = 2(N-1). \quad (8)$$

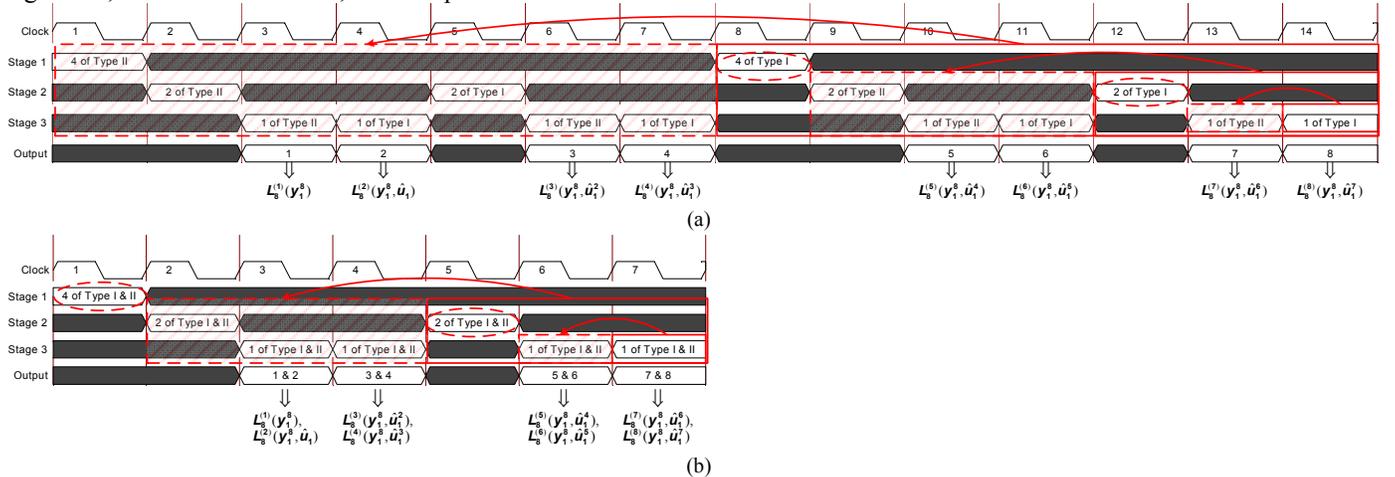

Figure 2: Conventional and look-ahead decoding time charts for polar codes with $N = 8$.

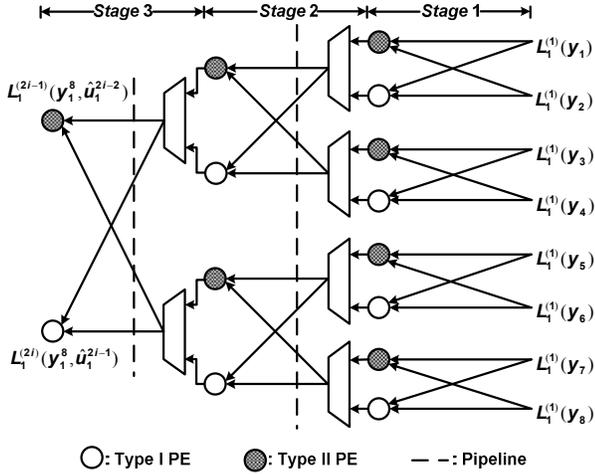

Figure 3: Pipelined decoder architectures of polar codes with length $N = 8$.

However, as mentioned previously the conventional decoding approach is not suitable for real-time communication systems for two reasons. First, in order to achieve the required performance, the code length $N$ is usually set as $2^{10}$-$2^{20}$. An immediate consequence is the latency of $2(N-1)$ clock cycles is too large. Second, it is apparent that during the whole decoding process the highest hardware utilization in a specific clock cycle is only 50% (*Clock cycle* 1). As the stage index increases, the hardware efficiency will go down as low as $1/N$ (*Clock cycle* $\log_2 N$), which can be lower than $2^{-10}$ for practical applications. Even for the pipelined tree architecture proposed by [3] in Fig. 3, the highest utilization is only 50% as well, which means half PEs are in idle state during each clock cycle.

This dilemma is introduced by the bottleneck of sequential decoding property of SC algorithm. It is noted that if both LLR inputs for Eq. (4) are available, there can be only two possible outputs, depending on what value $\hat{u}_{2i-1}$ will take. Therefore, for Type I PE, given both deterministic inputs, the look-ahead scheme only needs to pre-compute two output candidates, which can be selected by a multiplexer thereafter. For instance, shown in Fig. 1, all possible outputs of Type I PEs labeled by 8 in *Stage* 1 can be pre-calculated in *Clock cycle* 1. In other words, for *Stage* 1 the required computation in *Clock cycle* 8 can be incorporated into *Clock cycle* 1. In the similar way, for *Stage* 2 computation in *Clock cycle* 5 and 12 can be taken care of in *Clock cycle* 2 and 9, respectively. Calculation in *Clock cycle* 4, 7, 11, and 14 can be re-scheduled into *Clock cycle* 3, 6, 10, and 13 for *Stage* 3. As a result, only half clock cycles are required to implement the same decoding task with help of the proposed look-ahead schedule. For the 8-bit polar decoder example shown in Fig. 2 (b), all PEs at *Stage* 1 are activated during *Clock cycle* 1 because both deterministic LLR inputs for each PE are guaranteed by channel outputs. However, in *Clock cycle* 2, only PEs labeled with 2 or 5 can be activated, because they are the only ones with deterministic inputs. For PEs with labels of 9 or 12, their inputs are generated by Type I PEs in *Stage* 1, which have two possible values at this moment. In order to avoid error propagation caused by pre-computing to the next stage, those PEs stay idle during *Clock cycle* 2. Similar schemes apply to further decoding processes. It is clear that the required number of clock cycles can be halved to $N$-1. The time chart construction of the proposed scheme is given as follows:

| **Recursive Construction of Look-Ahead Time Chart** |
|---|
| 1: initializtion TC= *Null*; |
| 2: **for** $i = \log_2 N, i--,1$ **do** |
| 3:      $j = \log_2 N - i + 1$; |
| 4:      TC = $\{[j \text{ of Type I \& II}, TC], i\}$; |
| 5:      **if** $i = 1$ **then** |
| 6:          **break**; |
| 7:      **endif** |
| 8:      TC = [TC, TC]; |
| 9: **endfor** |
| 10: **output** TC. |

As indicated by *Step* 4, both types of PEs can work simultaneously in the same clock cycle, which not only shortens the decoding latency by 50% but also improves the hardware efficiency twice. Moreover, the proposed approach leads to a construction method in a recursive way. For clear understanding of the *Russian Doll*-like relationship between stages, the conventional and look-ahead construction processes have been pointed out with arrows in Fig. 2.

IV. ARCHITECTURES FOR LOOK-AHEAD DECODER

A. *Design of Type I PE*

According to the look-ahead scheme, Type I PE is in charge of pre-computing two possible outputs in parallel, which is in fact an adder-subtractor. Suppose $X$ and $Y$ are two operands, and $Z_{in}$ is the carried-in or borrowed-from bit. For the full adder the sum and carry-out bit are represented by $S$ and $C_{out}$. The difference and borrow-out produced by the full subtractor are denoted by $D$ and $B_{out}$. The logic equations are as follows:

$$S = X \oplus Y \oplus Z_{in}; \quad (9) \quad C_{out} = X \cdot Y + (X \oplus Y) \cdot Z_{in}; (10)$$

$$D = X \oplus Y \oplus Z_{in}; \quad (11) \quad B_{out} = \overline{X} \cdot Y + \overline{X \oplus Y} \cdot Z_{in}. \quad (12)$$

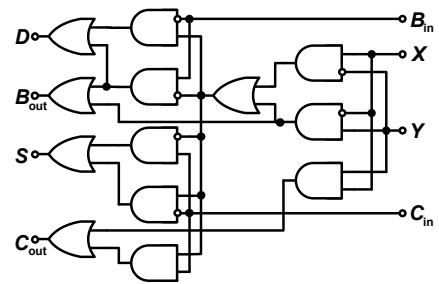

(a) 1-bit full adder-subtractor.

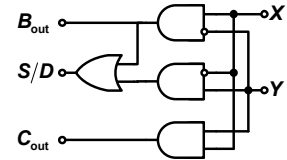

(b) 1-bit half adder-subtractor.

Figure 4: Proposed 1-bit adder-subtractor architectures.

Notice that $S$ and $D$ are actually the same, and $\overline{X}\cdot Y$ is an intermediate term of $X\oplus Y$. $\overline{(X\oplus Y)}\cdot Z_{in}$ is also a byproduct of $X\oplus Y\oplus Z_{in}$. The resulting gate-sharing scheme not only implements parallel processing but also reduces the hardware consumption. The gate-level structures of 1-bit full and half adder-subtractor are depicted in Fig. 4 (a) and (b), respectively. The proposed $q$-bit adder-subtractor, which is illustrated in Fig. 5, requires only less than 57% hardware compared with the conventional one while achieving the same performance.

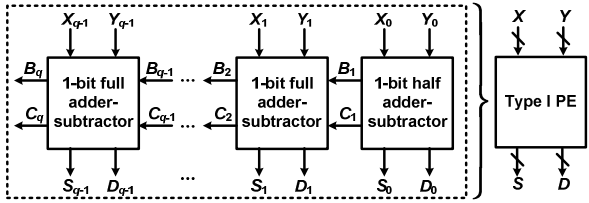

Figure 5: Proposed Type I PE architectures.

### B. Design of Type II PE

Type II PE with the min-sum algorithm is shown in Fig.6.

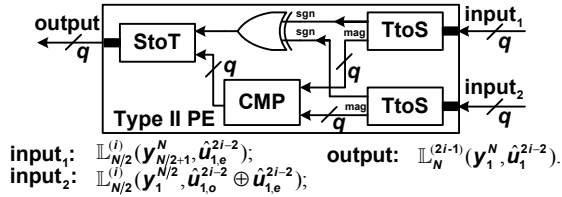

input$_1$: $\mathbb{L}_{N/2}^{(i)}(y_{N/2+1}^N, \hat{u}_{1,e}^{2i-2})$; output: $\mathbb{L}_N^{(2i-1)}(y_1^N, \hat{u}_1^{2i-2})$.
input$_2$: $\mathbb{L}_{N/2}^{(i)}(y_1^{N/2}, \hat{u}_{1,o}^{2i-2} \oplus \hat{u}_{1,e}^{2i-2})$;

Figure 6: Proposed architectures of Type II PE.

### C. Design of Merged PEs

Since the comparator in Type II PE is actually a $q$-bit subtractor, which is also employed by Type I PE, it is possible to merge Type I and Type II PEs together with the sub-structure sharing scheme. The detailed structure is as follows:

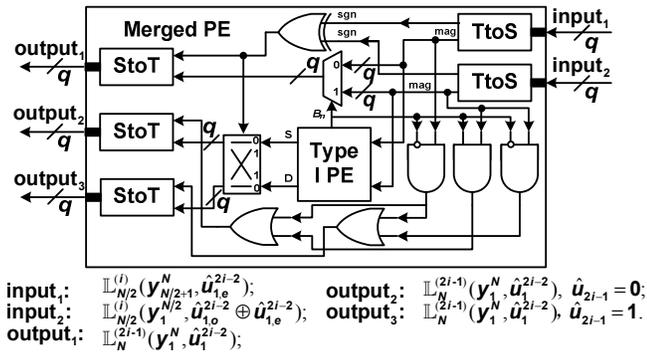

input$_1$: $\mathbb{L}_{N/2}^{(i)}(y_{N/2+1}^N, \hat{u}_{1,e}^{2i-2})$; output$_2$: $\mathbb{L}_N^{(2i-1)}(y_1^N, \hat{u}_1^{2i-2})$, $\hat{u}_{2i-1}=0$;
input$_2$: $\mathbb{L}_{N/2}^{(i)}(y_1^{N/2}, \hat{u}_{1,o}^{2i-2} \oplus \hat{u}_{1,e}^{2i-2})$; output$_3$: $\mathbb{L}_N^{(2i-1)}(y_1^N, \hat{u}_1^{2i-2})$, $\hat{u}_{2i-1}=1$.
output$_1$: $\mathbb{L}_N^{(2i-1)}(y_1^N, \hat{u}_1^{2i-2})$;

Figure 7: Proposed structure of the Merged PE.

### D. Input Generating Circuit for Type I PEs

As indicated in Eq. (4), except for $\mathbb{L}_{N/2}^{(i)}(y_1^{N/2}, \hat{u}_{1,o}^{2i-2} \oplus \hat{u}_{1,e}^{2i-2})$ and $\mathbb{L}_{N/2}^{(i)}(y_{N/2+1}^N, \hat{u}_{1,e}^{2i-2})$, a third input $\hat{u}_{2i-1}$ is also required by Type I PE. Moreover, for efficient execution of each Type I PE, the value of $\hat{u}_{2i-1}$ needs to be provided on-the-fly. However, even for the 8-bit decoder illustrated in Fig. 1, the complicated interleaving of odd and even indices makes the straightforward calculation of $\hat{u}_{2i-1}$ inconvenient. In order to solve this inherent problem, the input generating circuit (IGC) for Type I PEs is proposed in this section. Careful investigation has shown that it is possible to generate the required $\hat{u}_{2i-1}$ using the real FFT-like signal flow [4]. For instance, all $\hat{u}_{2i-1}$ for 8-bit polar decoder can be generated with the in-place procedure in Fig. 8.

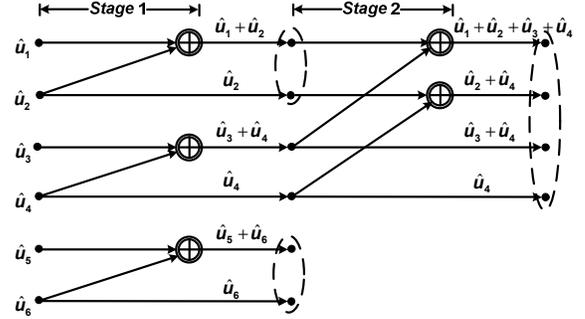

Figure 8: Flow graph of the proposed IGC.

The pipelined architecture of the flow graph is illustrated in Fig. 9, where $U_i$ denotes the unit which is consists of $i$ stage(s):

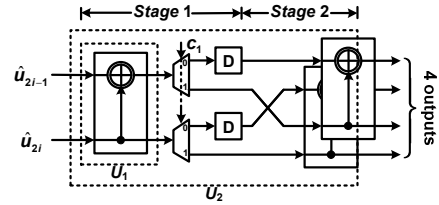

Figure 9: Pipelined architecture for the flow graph in Fig. 8.

In general, for $N$-bit length decoder, since the data structures of IGC are defined recursively for powers of 2, the pipelined architecture can be constructed with the recurrence relationship. The recursion for the general case is shown in Fig. 10, where module $U_n$ can be constructed based on module $U_{n-1}$ and $N/4$ extra XOR-pass elements. For efficient design, memory banks are employed instead of flip-flops. Here, $n = \log_2 N - 1$. Control signal $c_n$ can be obtained by down sampling $c_1$ by $n$.

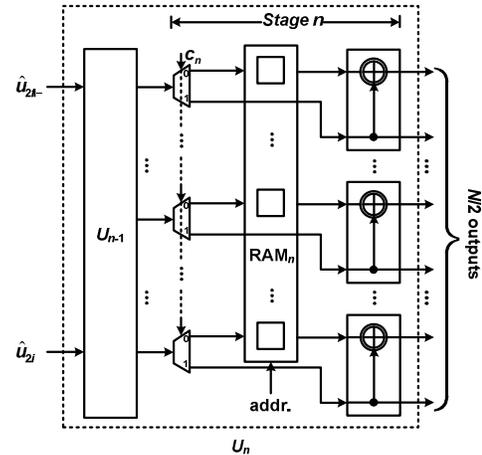

Figure 10: Recursive construction of $U_n$ based on $U_{n-1}$ using RAMs.

### E. Parallel Architecture of the Look-Ahead Decoder

Taking the advantage of the pre-stated blocks, the revised look-ahead decoder can be designed accordingly. Here we employ an 8-bit polar decoder as an example. However, it can be noticed that although hardware utilization for each active stage is 100%, other stages still remain idle at the same time.

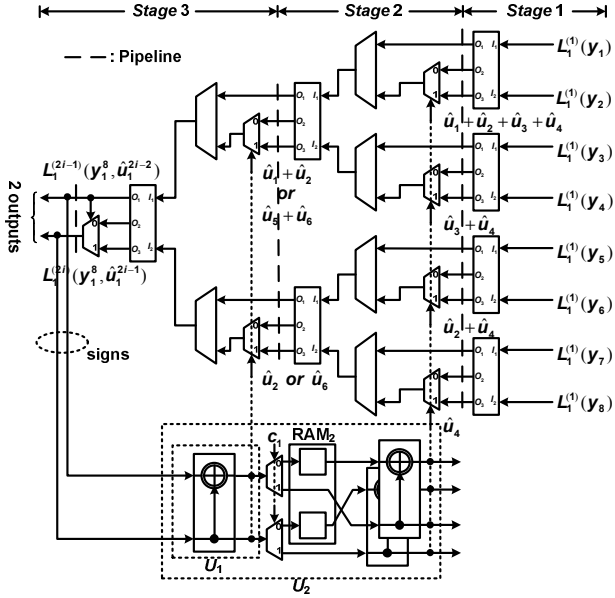

Figure 11: Pipelined decoder of look-ahead polar codes with $N = 8$.

Moreover, each codeword needs $N$-1 clock cycles to be properly decoded with the given approach, during which no new codeword could be processed. Therefore, a 2-parallel decoder is designed. Though an additional clock cycle is required as shown in Table I, for large $N$ it is negligible compared with the decoding latency. Also, twice higher throughput can be achieved by the decoder.

TABLE I NUMBER OF ACTIVE MERGED PEs IN EACH CLOCK CYCLE

| Input | Clock cycle | | | | | | | |
|---|---|---|---|---|---|---|---|---|
| | 1 | 2 | 3 | 4 | 5 | 6 | 7 | 8 |
| $C_1$ | 4 | | 2 | 1 | 1 | 2 | 1 | 1 |
| $C_2$ | | 4 | 2 | 1 | 1 | 2 | 1 | 1 |

## V. COMPARISON OF LATENCY AND HARDWARE

In this section, the proposed polar decoder is compared with the state-of-the-art reference. For the sake of fairness, both decoders have the same number of PEs. Since [3] failed to provide details of the $\hat{u}_s$ *computation block*, the counterpart of IGC, only comparison for the rest blocks is conducted.

TABLE II COMPARISON OF DIFFERENT POLAR DECODERS

| Different designs | | Proposed design | Line design [3] |
|---|---|---|---|
| Hardware consumption ($q$-bit quantization) | | | |
| # of Merged PEs | | $N/2$ | $N/2$ |
| 1 PE | XOR | $9q$ | $11q$-3 |
| | REG | 0 | 1 |
| | MUX | $6q$ | $5q$ |
| # of IGCs | | 2 | — |
| 1 IGC | XOR | $N/2$-1 | — |
| | RAM | $N/2$-2 | — |
| | MUX | $N/2$-2 | — |
| # of other REGs | | $q(9N/2+4)$ | $q(N-1)$ |
| # of other MUXs | | $q(N+2)$ | $3q(N/2-1)$ |
| Total | XOR[†] | $\sim 17qN/2$ | $\sim (19q-3)N/2$ |
| | REG | $\sim 9qN/2$ | $\sim (q+1/2)N$ |
| Decoding schedule | | | |
| Latency | | $N$ | $2(N-1)$ |
| Normalized throughput | | 2 | 1 |

[†]MUX is converted to XOR with the standard proposed in [5].

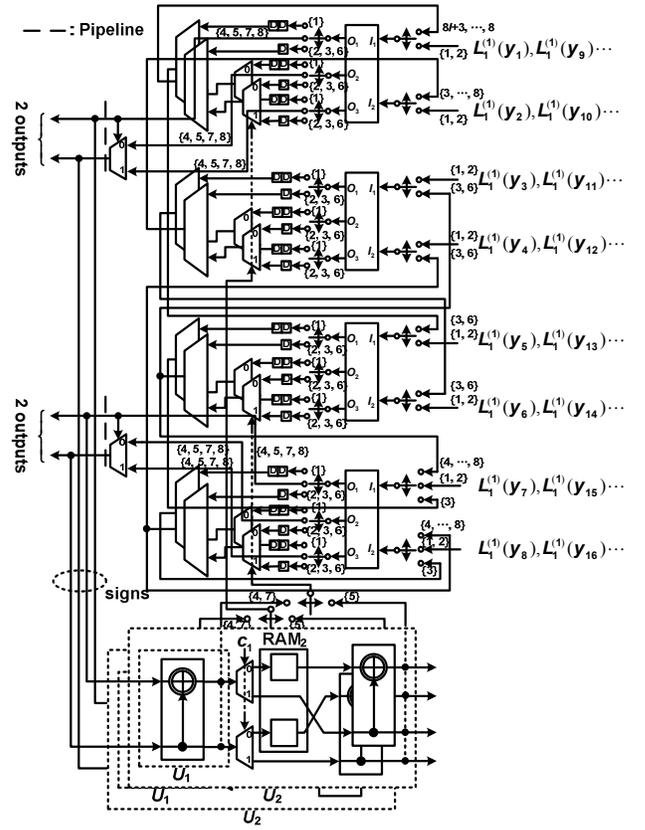

Figure 12: Parallel architectures for 8-bit polar decoder.

According to Table II, the given design only requires half latency as the reference does, while achieving twice higher throughput. And similar amount of hardware is required by the proposed one. Further discussion can show that the look-ahead approach is suitable for other SC decoders..

## VI. CONCLUSION

A novel look-ahead SC decoding schedule for polar codes is proposed in this paper, which can halve the decoding latency required by conventional approaches. For efficient hardware implementation issue, a *merged* PE and an IGC block are presented. Compared with its conventional counterpart, the parallel decoder example can halve the decoding latency and double the throughput with similar hardware consumption.

## REFERENCES

[1] E. Arikan, "Channel polarization: a method for constructing capacity-achieving codes for symmetric binary-input memoryless channels," *IEEE Trans. on Inf. Theory*, vol. 55, no. 7, pp. 3051-3073, July 2009.

[2] E. Arkan, "A performance comparison of polar codes and Reed-Muller codes," *IEEE Commun. Lett.*, vol. 12, no. 6, pp. 447-449, June 2008.

[3] C. Leroux, I. Tal, A. Vardy, and W. J. Gross, "Hardware architectures for successive cancellation decoding of polar codes," in *Proc. Int. Conf. Acoust., Speech, and Sig. Proc. (ICASSP)*, pp. 1665-1668, May 2011.

[4] M. Garrido, K. K. Parhi, and J. Grajal, "A pipelined FFT architecture for real-valued signals," *IEEE Trans. Circuits Syst. I: Reg. Papers*, vol. 56, no. 12, pp. 2634-2643, Dec. 2009.

[5] Xinmiao Zhang and Fang Cai, "Efficient Partial-Parallel Decoder Architecture for Quasi-Cyclic Nonbinary LDPC Codes," *IEEE Trans. Circuits Syst. I: Reg. Papers*, vol. 58, no. 2, pp. 402-414, Feb. 2011.